\documentclass[a4paper]{jpconf}
\pdfoutput=1
\usepackage{graphicx}
\usepackage{atlasphysics}
\usepackage{hyperref}
\begin{document}
\title{Experimental results with boosted top quarks in the final state}

\author{Johannes Erdmann for the ATLAS and CMS collaborations}

\address{Technische Universit\"at Dortmund, Fakult\"at Physik, Experimentalphysik IV, Otto-Hahn-Stra{\ss}e 4, 44227 Dortmund, Germany}

\ead{johannes.erdmann@cern.ch}

\newcommand{\seventev}{\ensuremath{\sqrt{s} = 7 \tev}}
\newcommand{\eighttev}{\ensuremath{\sqrt{s} = 8 \tev}}
\newcommand{\thirteentev}{\ensuremath{\sqrt{s} = 13 \tev}}

\begin{abstract}
  An overview of analyses using data at \seventev\ and $8\tev$ of proton-proton collisions
  at the LHC is presented. These analyses use boosted techniques to search
  for new phenomena involving top quarks and to measure the production of top
  quarks at high transverse momenta.
  Such techniques involve top-quark tagging algorithms, boson-tagging
  algorithms, and strategies for $b$-tagging and lepton identification in
  the environment where the top quark decay products are close to each other.
  The strategies are optimized for the different final states and for different ranges
  of the transverse momenta of the particles involved, improving on traditional resolved
  analysis strategies.
\end{abstract}

\section{Introduction}
With roughly $5\;\ifb$ of LHC proton-proton ($pp$)
collision data at \seventev\ and
roughly $20\;\ifb$ at $8\tev$ available for analysis, the ATLAS and CMS collaborations are facing the challenge of identifying top quarks in the {\it boosted} regime
with top-tagging techniques~\cite{boost2012} (top-tagging).
The traditional {\it resolved} strategy of identifying individual jets originating from the
decay of hadronically decaying top quarks fails for Lorentz boosts of the top quark for which
the decay products overlap in $\eta$--$\phi$ space.
With the use of larger jet radii, however, all decay products of
hadronically decaying top quark may be contained in one single jet (large-$R$ jets).
The substructure of such jets can be used to distinguish the three-prong decay of the
top quark from the background processes, in particular from jets originating from light
quarks or gluons.

Similar strategies are used for the identification of hadronic decays of $W$, $Z$ and Higgs
bosons (boson-tagging): $W \rightarrow q\bar{q}^\prime$, $Z \rightarrow q\bar{q}$ and $H \rightarrow b\bar{b}$.
The identification of jets originating from $b$-quarks, is more challenging in the boosted
regime, because of nearby activity in the tracking detector and the high transverse
momentum of $b$-quarks.
The application of $b$-tagging to small-$R$ subjets of the large-$R$ jet (subjet $b$-tagging), however,
improves the performance of the top-tagging algorithms, and allows for multiple
$b$-tags inside a single large-$R$ jet as used in the identification of hadronically decaying Higgs bosons.
Moreover, lepton identification is more challenging in the boosted regime, because
nearby activity renders traditional isolation criteria inefficient.
Hence, analyses using leptons often make use of modified lepton-isolation criteria.
The top-tagging, boson-tagging and $b$-tagging strategies used by the ATLAS and CMS collaborations
in the boosted regime are described in several documents~\cite{Ctoptag,CWtag,Cbtag,Atoptag,AWtag},
where also predictions from Monte Carlo (MC) simulations are compared to $pp$ collision data.

In this document, analyses in the all-hadronic final state are discussed separately from
analyses in final states with at least one lepton, because they
require techniques with high suppression of the large QCD multijet background.
In the analyses discussed in this document,
the various background contributions are estimated using combinations of MC
simulations and control regions.
Single top, $\ttbar+W/Z$ and $WW/WZ/ZZ$ production are estimated using MC
simulation.
$\ttbar$ and $W/Z+{\rm jets}$ production are estimated using MC simulation, but also approaches with estimates from control regions are used in some analyses.
Contributions from QCD multijet background are estimated from data from background enriched control regions.
All analyses discussed use the full \eighttev\ data set, if not stated differently.

\section{All-hadronic final states}
This section reports on analyses in the all-hadronic final state, which make use of
boosted techniques in the search for supersymmetric partners of the top quark,
in the search for a vector-like partner of the top quark, and in the search for $\ttbar$
and $tb$ resonances, such as $Z^\prime$ and $W^\prime$ bosons.
Different top-tagging techniques are used, each taking into account the topology of
the respective signal processes and the relevant range in top-quark
transverse momentum.

\subsection{Search for supersymmetry}
Both, ATLAS and CMS have searched for supersymmetric partners of the top quark, {\it stop}
quarks $\tilde{t}$, in the all-hadronic decay channel:
$\tilde{t}\bar{\tilde{t}} \rightarrow \ttbar + 2 {\rm LSP} \rightarrow {\rm jets} + \met$,
where the missing transverse momentum arises from the lightest supersymmetric particle (LSP)
escaping detection.
The boost of the top quarks is driven by the mass difference of the stop quark and the
LSP, which only leads to moderate boosts for the relevant supersymmetry models considered.
Both analyses use strategies targeting the partially boosted topology.
In the ATLAS analysis~\cite{AsusyHad} resolved selections with large \met\ and at least two $b$-tags
are used.
These are complemented by a selection using
two large-$R$ jet radii of $R=1.2$ and $R=0.8$ to identify top-quark and $W$-candidates.
The large-$R$ jets are built from small-$R$ jets rather than calorimeter clusters in order to
preselect the relevant structure of the event.
Kinematic requirements on the top-quark and $W$-boson candidates are imposed to enhance
the sensitivity of the analysis, allowing in particular for the identification of asymmetric top-quark pairs,
where only the higher \pt\ candidate is reconstructed.
In the CMS analysis~\cite{CsusyHad}, also a small-$R$ jet preselection with large \met\ and at least
one $b$-tag is used.
Small-$R$ jets are then used to reconstruct one top-quark candidate with HEPTopTagger-like~\cite{HTT}
selection requirements.
The second top quark is allowed to be only partially reconstructed using the remaining
jets in the event by investigating three-jet and two-jet masses and the $b$-tagging
properties of these jets.
The data samples used in the analysis is then divided into different categories based on the number of $b$-tags
and the value of \met\ present in the event.
No evidence for supersymmetric partners of the top quark is found, and both analyses
set upper limits at 95\% confidence level (CL) in the 2D plane spanned by the stop-quark
mass and the LSP mass, reaching up to stop quark masses of 645~\gev.

\subsection{Search for vector-like quarks}
CMS has searched for pair production of a vector-like top-quark partner with an
electric charge of $+\frac{2}{3}$, $T$, in the decay channel
$T\bar{T} \rightarrow tH\bar{t}H \rightarrow {\rm jets}$~\cite{CthHad}.
Cambridge-Aachen (C/A) jets with a radius parameter of $R=1.5$ are used to account for the moderate
boost of the top quarks and Higgs bosons given the large mass in the final state of this analysis.
In order to suppress the dominant QCD multijet background, 
top-quark candidates are identified by jets which are tagged with the HEPTopTagger and at
least one $b$-tagged HEPTopTagger subjet.
Higgs-boson candidates are identified by requiring two subjets to be $b$-tagged and their
invariant two-jet mass to be larger than $60\gev$.
At least one top-quark and one Higgs-boson candidate are required in the event selection,
and the events are further categorized by the number of Higgs-boson tags.
The scalar sum of all jet transverse momenta is used as the final discriminant together with
the invariant two-jet mass of the Higgs-boson candidate.
No excess over the Standard Model prediction is found and 95\% CL limits are set on the
cross section times branching ratio (BR) for $T$ pair production, excluding models up
to $T$ masses of 747~\gev.

\subsection{Search for $\ttbar$ resonances}
ATLAS and CMS have searched for $\ttbar$ resonances in the all-hadronic decay
channel.
In the ATLAS analysis~\cite{AttHad}, which uses the full \seventev\ data set, two different
top-tagging strategies are followed:
the HEPTopTagger is used to ensure sensitivity to low invariant $\ttbar$ masses with
a C/A jet radius of $R=1.5$ and a minimum jet \pt\ of $200\gev$.
The Template Overlap Method~\cite{TOM1} is used for better sensitivity at high invariant $\ttbar$ masses.
An anti-$k_t$ jet radius of $R=1.0$ is used with a minimum jet \pt\ of $500\gev$ and $450\gev$
for the leading and subleading jet, respectively.
For both top-tagging techniques, two top-quark candidates have to be present in the event,
and a $b$-tagged small-$R$ jet has to be matched to each of them.
The CMS analysis~\cite{CttAll} uses full \eighttev\ data set and is optimized for higher transverse momenta.
As before, two top-quark candidates are required.
The CMS tagger, which is based on the JHU tagger~\cite{JHU}, is used with C/A jets with a radius parameter of $R=0.8$ and a minimum
\pt\ of $400\gev$.
No $b$-tagging requirements are imposed, but only angular requirements are made, which
require the two top-quark candidates to be back-to-back in azimuth.
No evidence for deviations from the Standard Model prediction is found and 95\% CL
limits are set on several benchmark model cross sections times BR to $\ttbar$,
reaching up to resonance masses of 2.1~\tev for a $Z^\prime$ model in combination with
the analysis in the leptonic decay channels.

\subsection{Search for $tb$ resonances}
ATLAS and CMS have searched for heavy partners of the $W$ boson, $W^\prime$ bosons,
which decay to a top quark and a $b$-quark using the hadronic decay channel
of the top quark.
Both analyses are optimized for the high-mass region, where the sensitivity from searches
using the leptonic decay of the top quark is lower with standard lepton isolation requirements, because the lepton from the
top-quark decay overlaps with the $b$-quark from the top-quark decay.
Both analyses require one top-quark and one $b$-quark candidate to be identified.
In the ATLAS analysis~\cite{AtbHad}, top-quark candidates are identified using a combination of
requirements on the one-to-two splitting scale~\cite{d12} and $N$-subjettiness~\cite{Nsub1} ratios based on
anti-$k_t$ jets with a radius parameter of $R=1.0$ and $\pt > 350\gev$.
Also, a $b$-tagged small-$R$ jet has to be geometrically matched to the large-$R$ jet.
In the CMS analysis~\cite{CtbHad}, a combination of the CMS-tagger, a $N$-subjettiness ratio and
a subjet $b$-tagging requirement is used to identify top-quark candidates based on
C/A jets with $R=0.8$ and $\pt > 450\gev$.
In the ATLAS analysis, one $b$-tagged small-$R$ jet with $\pt > 350\gev$ is required.
Also in the CMS analysis, the $b$-quark candidate has to be $b$-tagged, but the same jet algorithm as for the top-quark candidate is used.
Additionally the mass of the $b$-quark candidate jet is required to be smaller than
$70\gev$ in order to suppress background from \ttbar\ production.
No deviations from the Standard Model expectation are found and 95\% CL limits are set
on the cross section times BR in a left-handed $W^\prime$ model as well as in a right-handed
$W^\prime$ model. Limits are also set in the 2D coupling-mass plane and on the relative
admixture of left- and right-handed $W^\prime$ bosons in the case of the ATLAS and CMS
analysis, respectively.
$W^\prime$ models with masses up to 2.15~\tev are excluded.

\section{Final states with one or more leptons}
In the presence of at least one lepton in the final state, electron and
muon identification can be used to significantly reduce the background contribution
from QCD multijet production.
Hence, different strategies are used to identify boosted objects, and choices with higher
efficiency are made, for example for the top-tagger used.
However, the reduced efficiency of traditional lepton isolation criteria needs to be
addressed by modified isolation requirements, for example by using
an isolation cone with decreasing size for increasing \pt\ 
(mini-isolation)~\cite{miniiso}: $\Delta R = x \GeV / \pt$.

In this section, analyses search for supersymmetric partners of the top quark, vector-like partners of the $b$- and top quark,
and for $\ttbar$ resonances are described, as well as a measurement of the differential $\ttbar$ production cross section
at high top-quark \pt.

\subsection{Search for supersymmetry}
Similarly to the search for supersymmetric partners in the all-hadronic top quark decay
mode discussed above, ATLAS also makes use of boosted techniques in a search using
the $\ttbar$ single lepton decay channel~\cite{AsusyLep}.
In this analysis, 15 signal categories are defined, which are sensitive to different
kinematic configurations of stop-quark pair production.
One of these categories is optimized for sensitivity to a signal with boosted top quarks.
Given that the \pt\ of top quarks only reaches moderate values in the scenarios considered, a loose top-tagging
requirement is used with anti-$k_t$ $R=1.0$ jets with $\pt > 270 \gev$ and a mass
requirement of at least $70\gev$.
These criteria also allow for a partial containment of the top quark decay products in
the large-$R$ jet.
It was shown that the boosted selection improves over the resolved selections
in the case of large stop-quark masses and small LSP masses, i.e. in the case of relatively
large top-quark boost.
No evidence for stop-quark pair production is found and 95\% CL limits are set in the 2D
plane spanned by the stop-quark and LSP masses, and stop masses up to 640~\gev\ are
excluded.

\subsection{Search for vector-like quarks}
CMS has searched for the pair production of vector-like $b$-quark partners
$b^\prime \to tW/bZ/bH$~\cite{Cbprime},
charge $+\frac{2}{3}$ top-quark partners
$T \to bW/tZ/tH$~\cite{Ctprime},
and charge $+\frac{5}{3}$ top-quark partners
$T_{5/3} \to tW$~\cite{Ct53}.
These analyses make use of C/A $R=0.8$ jets and in categorize events according to the
number of observed top-quark or boson tags.
The CMS top-tagger is used as well as jet mass criteria and mass drop~\cite{massdrop} criteria in order to identify
jets originating from $W$ and $Z$ bosons.
No evidence for vector-like quark production is found and 95\% CL limits are set on the
pair production cross sections times BR, and $b^\prime$, $T$ and $T_{5/3}$ vector-like
quarks are excluded up to masses of 732, 782 and 800~\gev, respectively.

\subsection{Search for $\ttbar$ resonances}
CMS and ATLAS have searched for resonant $\ttbar$ production in the single lepton
decay channel~\cite{CttAll,AttLep}, where the ATLAS analysis uses $14~\ifb$ of
data taken at \eighttev, and the CMS analysis uses the full \eighttev\ data set.
Both analyses use a resolved selection and a complementary, boosted selection optimized
for sensitivity at high invariant $\ttbar$ masses.
A $\chi^2$ algorithm for $\ttbar$ reconstruction is used in the CMS analysis which exploits
the $\ttbar$ decay topology to reject background contributions.
The $\chi^2$ algorithm is modified for the boosted selection to ensure high
sensitivity to large resonance masses.
Electrons and muons do not need to fulfill isolation criteria, and a high lepton
identification efficiency is retained even for large top-quark transverse momenta.
In order to suppress background contributions, in particular from $W$+jets production,
a hadronic top-quark candidate is reconstructed using anti-$k_t$ jets with $R=1.0$,
$\pt > 300 \gev$ and a loose top-tagging requirement based on the mass ($m > 100 \gev$)
of the jet
and its one-to-two splitting scale ($\sqrt{d_{12}} > 40\gev$).
In order to ensure high lepton-identification efficiencies up to large top quark \pt, mini-isolation is used.
No evidence for deviations from the Standard Model prediction is found and 95\% CL
limits are set on several benchmark model cross sections times BR to $\ttbar$, and
a $Z^\prime$ model is excluded with masses up to 1.8~\tev.

\subsection{Differential $\ttbar$ production cross section}
Using the boosted selection of the $\ttbar$ resonance search in the single lepton
channel, ATLAS has measured the $\ttbar$ production cross section as a function of
the \pt\ of the hadronically decaying top quark reconstructed using boosted
techniques~\cite{Adiff}.
The use of the large-$R$ radius jets and the top-tagging requirements allow for a measurement
of the cross section in the region of \pt\ as high as 800--1200\gev.
The differential cross section is unfolded to the particle level and is compared to various
predictions from leading-order multileg generators and next-to-leading order generators.
The MC simulations investigated fail to describe the data, in particular at high transverse
momentum of the top quark, and predict a higher cross section at high \pt.

\section{Summary and conclusions}
An overview of analyses using LHC \seventev\ and $8\tev$ $pp$ collision data was presented.
The ATLAS and CMS collaborations have used boosted techniques for top identification,
$W$, $Z$ and Higgs boson identification, $b$-tagging strategies and modified lepton
identification criteria to search for new phenomena beyond the Standard Model.
The use of the boosted techniques improves the sensitivity of these searches compared to
traditional resolved analysis strategies.
The ATLAS collaboration has performed a measurement of the differential $\ttbar$ production
cross section in the single lepton decay channel, as a function of the transverse momentum
of the hadronic top quark.
The use of boosted techniques allows to increase the reach of this measurement to higher
transverse momenta compared to an analysis considering only the resolved topology.
In the light of the upcoming Run-2 of the LHC at \thirteentev, boosted techniques will gain even
more importance for searches for new phenomena at the $\tev$ scale, but also for
measurements of top quarks with high transverse momenta.
The experience gained with boosted techniques in Run-1 analyses is a promising basis for
the upcoming challenges.

\ack{The author would like to thank Emanuele Usai for his help on the CMS analyses.}

\end{document}